\documentclass[12pt,preprint]{aastex}

\slugcomment{To appear in The Astrophysical Journal}

\begin{document}


\title{Nucleosynthesis in the early Galaxy}


\author{F.~Montes\altaffilmark{1,10}, T.C.~Beers\altaffilmark{2,3},
J.~Cowan\altaffilmark{4}, T.~Elliot\altaffilmark{2,5,6},
K.~Farouqi\altaffilmark{7}, R.~Gallino\altaffilmark{8},
M.~Heil\altaffilmark{1}, K.-L.~Kratz\altaffilmark{7,9},
B.~Pfeiffer\altaffilmark{7}, M.~Pignatari\altaffilmark{8},
H.~Schatz\altaffilmark{2,5,6}}


\altaffiltext{1}{Gesellschaft f{\"u}r Schwerionenforschung, D-64220
Darmstadt, Germany} \altaffiltext{2}{Joint Institute for Nuclear
Astrophysics, http://www.jinaweb.org, USA} \altaffiltext{3}{Dept. of
Physics and Astronomy and Center for the Study of Cosmic Evolution
(CSCE), Michigan State University, E. Lansing, MI 48824, USA}
\altaffiltext{4}{Homer L. Dodge Dept. of Physics and Astronomy,
University of Oklahoma, Norman, OK 73019 USA} \altaffiltext{5}{Dept.
of Physics and Astronomy, Michigan State University, E. Lansing, MI
48824, USA} \altaffiltext{6}{National Superconducting Cyclotron
Laboratory, Michigan State University, E. Lansing, MI 48824, USA}
\altaffiltext{7}{Virtuelles Institut f{\"u}r Struktur der Kerne und
Nukleare Astrophysik (VISTARS), D-55128 Mainz, Germany}
\altaffiltext{8}{Dipartimento di Fisica Generale, Universita' di
Torino, Via P. Giuria 1, I-10125 Torino, Italy}
\altaffiltext{9}{Max-Planck-Institut f{\"u}r Chemie,
Otto-Hahn-Institut, Joh.-J.-Becherweg 27, D-55128 Mainz, Germany}
\altaffiltext{10}{Present address: NSCL/MSU, East Lansing, MI-48824,
US}


\begin{abstract}
Recent observations of r-process-enriched metal-poor star abundances
reveal a non-uniform abundance pattern for elements $Z\leq47$. Based
on non-correlation trends between elemental abundances as a function
of Eu-richness in a large sample of metal-poor stars,
it is shown that the mixing of a consistent and robust 
light element primary process (LEPP) and the r-process pattern found
 in r-II metal-poor stars explains such apparent non-uniformity.
 Furthermore, we derive the abundance pattern of the LEPP from observation
 and show that it is consistent with a missing component in the solar
 abundances when using a recent s-process model. As the astrophysical
 site of the LEPP is not known, we explore the possibility of a neutron
 capture process within a site-independent approach. It is suggested that
 scenarios with neutron densities $n_{n}\leq10^{13}$ $cm^{-3}$ or in the
 range $n_{n}\geq10^{24}$ $cm^{-3}$ best explain the observations.
\end{abstract}




\section{Introduction}
The r-process is responsible for the origin of about half of the
heavy isotopes beyond the iron group in nature, yet its site is
still not determined with certainty \citep{CTT91,TCP02,cowan06}.
The r-process involves extremely unstable nuclei and in order for
neutron captures to overcome the correspondingly short beta decay
rates, typical conditions with neutron densities in excess of
$10^{20}$ cm$^{-3}$ and a process duration of less than $\approx$5 s
are required \citep{kratz93}. R-process models have had difficulties
in obtaining such conditions in realistic astrophysical
environments. One promising scenario and one of the most studied is
the neutrino-driven wind off a proto-neutron star in core collapse
supernovae \citep{WWH92,TWJ94,Wan01,Tho01,Far06}. Alternative
scenarios include neutron star mergers \citep{FRT99,ROS99,GDJ05},
jets in core collapse supernovae \citep{Cam01}, and gamma ray bursts
\citep{SuM05}. The s-process is responsible for creating roughly the
other half of the isotopes beyond the iron group, while the
p-process has significant contributions only on the relatively rare
isotopes on the proton-rich side of the nuclei chart. It has also
been recognized that to correctly reproduce the solar system
s-process abundances at least two different components are required
\citep{kaeppeler82}; the weak s-process component responsible for
creating s-isotopes with  $A\leq90$ \citep{raiteri92,PIG06}, and the
main s-process responsible for the heavier s-isotopes
\citep{Arlandini99,Travaglio04,CRI06}. A third s-process component,
called strong s-process, was envisaged by \citet{clayton67} at the
termination of the s-process, to account essentially for about 50\%
of solar $^{208}$Pb. This component was recently recognized as the
outcome of AGB stars of low metallicity \citep{travaglio01}.

The solar system r-process abundances are often inferred by using
the calculated s- and p- abundances and subtracting them from the
observed solar system abundances. If there is an additional
nucleosynthesis process creating only small amounts of residual
abundances, its contribution may be ``hidden" in the such defined
solar system r-process abundances.

In addition to the solar system abundance distribution, observations
of elemental abundances in unevolved metal-poor halo stars can
provide important clues about nucleosynthesis events in the early
Galaxy. These stars are old and preserve in their photospheres the
abundance composition at the location and time of their formation.
In particular a sub-class of extremely metal-poor
([Fe/H]$\approx-$3) but Eu enhanced stars ([Eu/Fe]$>$0.5) exhibit
what appears to be a pure r-process abundance pattern for the heavy
r-process elements $Z\geq56$ and $Z<83$. This pattern is remarkably
stable from star to star and in excellent agreement with the
contribution of the r-process to the solar abundances. A few of
these stars have been found to date, with CS~22892-052
\citep{sneden03} being the prime example. Since they are thought to
exhibit the abundance pattern produced by a single or at most a few
r-process events in the early Galaxy, the stability of the observed
abundance pattern and the good agreement with the solar system
r-process contribution imply that r-process events generate a
universal abundance distribution. The universality of the abundance
pattern of the heavy r-process elements seems not to extend to the
actinides Th and U, where some star to star scatter has been found
in some cases \citep{hill02,Gor02,Sch02,honda04}.

While the r-process abundance pattern for $56\leq Z<83$ is stable
from star-to-star, the overall level of enrichment with respect to
iron (for example [Eu/Fe]) shows a very large star to star scatter.
This implies that very metal-poor halo stars sample a largely
unmixed early Galaxy and that the r-process occurs in at most
2-10$\%$ of iron producing core collapse supernovae \citep{TCP02}
(assuming core collapse supernovae are the site of the r-process).

However, this simple picture breaks down for the lighter neutron
capture elements with $Z\leq47$. The abundances of these elements
measured in the strongly Eu-enhanced stars once normalized to Eu do
not agree entirely with the solar system r-process residual pattern.
In particular the abundances of Y, Mo, Rh, Pd and Ag are
consistently below the solar system values. This either indicates that the r-process in this region is not robust and depends on the astrophysical condition or metallicity, or that the r-process observed in these stars is not the only nucleosynthesis process leading to the abundance pattern obtained by subtracting the s- and p- processes from the solar abundances \citep{pfeiffer01a,pfeiffer01b}. There is
additional evidence for such a second process being present in the
early Galaxy: \citet{wasserburg96} and \citet{qian00a} first proposed
the existence of two different r-process sites or components based
on meteoritic evidence of live light r-process $^{129}$I
(T$_{1/2}$=15.7 Myr) in the early solar system compared to the heavy
r-process isotopes such as $^{182}$Hf (T$_{1/2}$=8.9 Myr,
\citet{vockenhuber04}). \citet{qian00b,qian01,qian03} 
also proposed that two r-processes together with a ``prompt" nucleosynthesis
 contribution could explain the metal poor star abundance observations
 available at the time. However, it has been pointed
out that the proposal of different r-process sites for the isotopes
in the second r-process peak from I to Te
(including $^{129}$I) and for isotopes of Ba and beyond (including $^{182}$Hf) are
difficult to reconcile with r-process models and the known nuclear
physics at the $N=82$ shell gap \citep{kratz06}. \citet{mcwilliam98},
\citet{burris00}, \citet{norris01}, \citet{johnson02},
\citet{lambert02}, \citet{TCP02}, \citet{honda04} and
\citet{barklem05} reported the observation of a large scatter in
[Sr/Ba] in low-metallicity stars. This has been interpreted as further 
evidence for a second, independent process
that produces Sr but little or no Ba at low metallicities.
\citet{Travaglio04} demonstrated that the same is true for Y, and
Zr, and they postulated a LEPP - Light Element Primary Process
producing Sr, Y, and Zr, but little Eu and Ba. \citet{TCP02} pointed
out that there is in fact a non-correlation of [Sr/Ba] and [Ba/Fe]
in some stars showing very large [Sr/Ba] ratios but little Ba.
Again, this can be explained by the presence of a second process
producing mainly Sr that happens to dominate the composition in such
stars. A similar non-correlation was found by \citet{kaori06} for a
few stars in the globular cluster M15. \citet{aoki05} came to a
similar conclusion based on trends in the behavior of Y and Zr as a
function of Eu.

We show here in $\S$~\ref{abundanceclues} that such an
non-correlation can be found in all metal-poor, r-process-enriched
stars, and for additional light elements beyond Sr, Y, and Zr. This
includes previously noted ``anomalies" such as the observed
abundances in HD~122563 \citep{honda06} that cannot be fit by an r-
or an s-process alone. Also included is the abundance pattern in the
moderately r-element enhanced star HD~221170 \citep{ivans06} which
does not exhibit the underabundance of light r-process elements with
respect to the solar r-process contribution. We show that these
``anomalies" are in fact part of a general trend that is consistent
with a mixture of two processes in metal-poor stars, an r-process
and a LEPP process exhibiting rather stable abundance patterns,
which are mixed in varying proportions. In $\S$~\ref{modelsection}
we then analyze the features of the newly derived LEPP abundance
pattern in terms of a neutron capture process model to determine the
astrophysical conditions required for this new process. Conclusions
are presented in $\S$~\ref{conclusions}.

\section{Abundance clues}
\label{abundanceclues}
\begin{figure}
\includegraphics[scale=0.70]{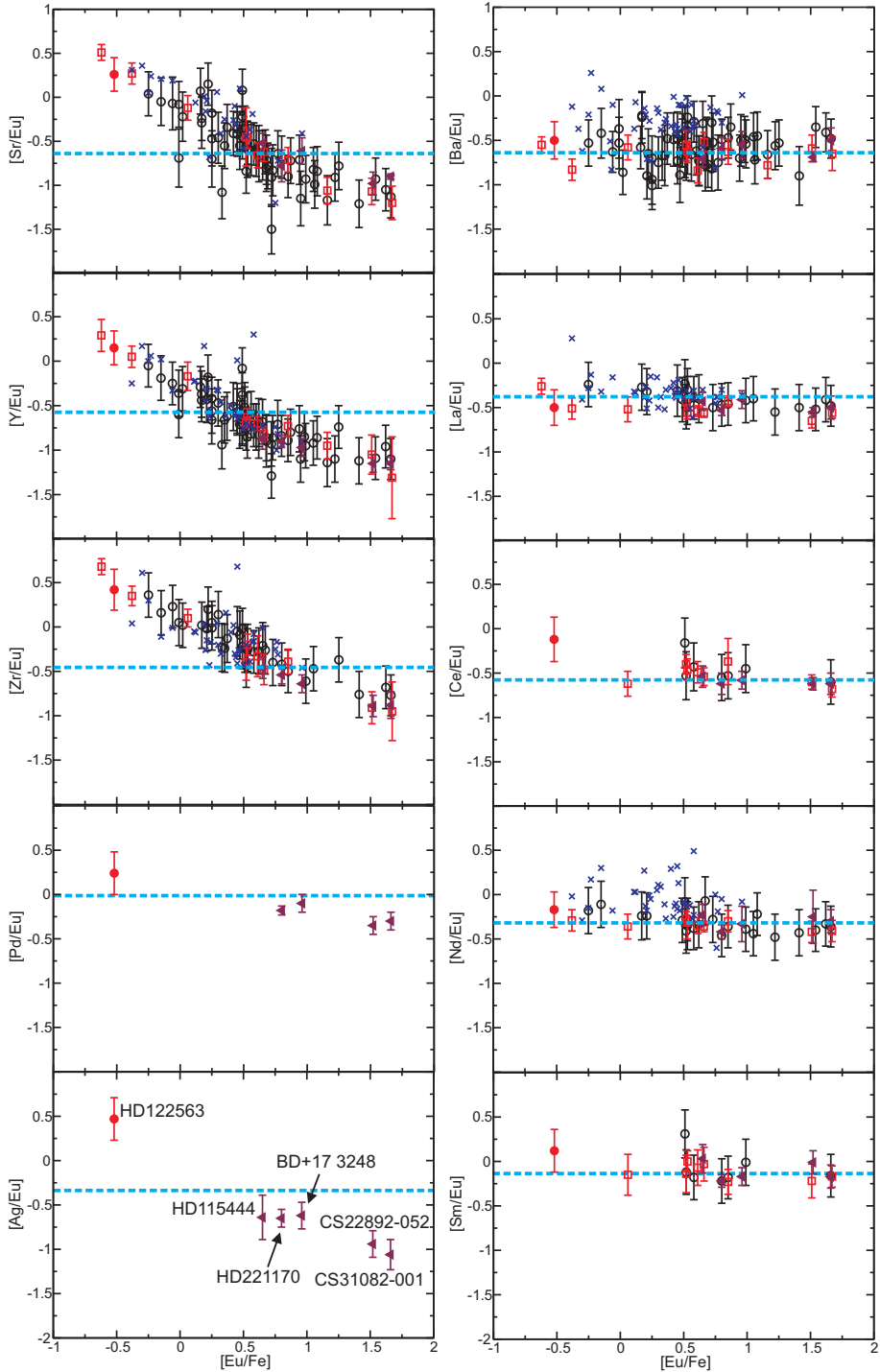}
\caption{Abundance ratio of metal-poor stars as a function of
[Eu/Fe]. Only stars with [Ba/Eu]$<$0 (except HD~122563) and
[Fe/H]$<-$1 are shown. Abundances represented by crosses were taken
from \cite{burris00}, open squares from \cite{honda04}, open circles
from \cite{christlieb05,barklem05} and filled circles from
\cite{honda06}. Dashed lines are the respective r-process ratios.}
\label{ratios}
\end{figure}

The observed elemental abundances of metal-poor ([Fe/H]$<-$1) and
r-process enriched ([Ba/Eu]$<$0) stars are shown in
Fig.~\ref{ratios}
\citep{burris00,honda04,christlieb05,barklem05,honda06}. All
abundances are normalized to Eu, which is predominantly an r-process
element ($\approx$95$\%$ of the total solar abundance). The heavy
r-process elements with $Z\geq56$ exhibit, within the observational
errors, a constant ratio with respect to Eu, independent of the Eu
enrichment of the star. This ratio is consistent with the element
ratio of the r-process contribution to the solar system. This
indicates that the heavy elements are produced by the r-process that
produces a universal abundance pattern with fixed element ratios
consistent with the solar r-process abundance pattern. This is the
same conclusion that has been drawn from the abundance pattern of
the few very metal-poor, strongly r-process enhanced stars where a
large range of elemental abundance have been determined (see for
example the reviews by \citet{TCP02,cowan06}).

It is also obvious from Fig.~\ref{ratios} that the lighter r-process
elements behave very differently. Clearly, the [Y/Eu], [Sr/Eu] and
[Zr/Eu] ratios are not constant but show a non-correlation with the
Eu enrichment. This non-correlation indicates that a different
process (the LEPP), which does not create substantial amounts of
heavier r-process elements such as Eu, has contributed significantly
to the abundances of Sr, Y, and Zr. \citet{kaori06} studying 6 giants in the globular cluster
M15 observed a similar relation for Y and Zr with respect to Eu and
came to the same conclusion, but with very low statistics. We
confirm their result with a much larger sample and show that this is
true for all r-process enhanced, metal-poor stars. While observations require this process to be different in the sense that it operates not always simultaneously with the process that produces the r-process abundances observed in highly r-process-enhanced stars (r-II, with
[Eu/Fe]$>$1.0 and [Ba/Eu]$<$0.0 , according to \citet{beers05}), it is still possible that both processes occur in the same astrophysical object.

The [X$_i$/Eu] versus [Eu/Fe] slopes for the light r-process elements in Fig.~\ref{ratios} are at least for not too large [Eu/Fe] roughly consistent with -1. This is a consequence of the correlation of the light r-process elements with Fe instead of Eu. As \citet{Travaglio04} have shown for Sr, Y, and Zr, [X$_i$/Fe] is roughly constant and shows a rather small scatter as a function of metallicity [Fe/H]. Because of [X$_i$/Eu]$=$[X$_i$/Fe]$-$[Eu/Fe], this results in a -1 slope in [X$_i$/Eu] versus [Eu/Fe].

An interesting question is the behavior of other light r-process
elements below Ba. Ag would be a good indicator for the r-process as
$\approx$80-86$\%$
\citep{Arlandini99,burris00,Travaglio04}\footnotemark[11] of the
total solar abundance is produced in the r-process. Unfortunately,
only metal-poor stars CS~22892-052 \citep{sneden03}, HD~155444
\citep{westin00,sneden06}, BD~+17$^{o}$3248 \citep{cowan02},
CS~31082-001 \citep{hill02}, HD~221170 and HD 122563 have published
abundance yields of Nb, Mo, Ru, Rh, Pd and Ag\footnotemark[12]. Nevertheless,
although the statistics are low for Pd and Ag, the elemental
abundances shown in Fig.~\ref{ratios} are more consistent with an Eu
non-correlation as observed for Sr, Y, and Zr, than with the
constant ratio exhibited by the heavier r-process elements. We must
caution, however, that the observed Ag abundances in these
metal-poor stars may have some uncertainties. The atomic data for
this element are well established and the Solar System abundance
also appears well determined, but it is not clear whether non-LTE
effects could have affected the abundance analysis since the Ag abundances are based upon (low-lying) neutral transitions as opposed to ion transitions.

Within their
very low statistics (2-4 data points) the abundances of Nb, Mo, and
Rh  are also consistent with the trend exhibited by Sr, Y, Zr, Pd
and Ag. A possible exception  is Ru, which shows a flat trend, but
it is not possible to make definite conclusions based on just 3 data
points. Also, Ru has less established atomic data for the lines
analyzed. We therefore conclude that the LEPP not only produces Sr,
Y, and Zr, but most likely all light elements between Sr and Ag
observed in very low metallicity stars.

\footnotetext[11]{A reanalysis of the calculations in
\citet{Travaglio04} indicates some modification in the  values for
the Ag entries in their table 5. The corrected $s$ fraction is 14\%
and the $r$ residual is then 86\%.}

\footnotetext[12]{Recent high resolution spectroscopy of the extremely metal-poor star HD~88609 was recently reported in \citet{honda07} after the manuscript was completed and it is not included in the discussion. Its elemental distribution is very similar to the one observed in HD~122563.}

Fig.~\ref{ratios} also shows the underproduction of Sr, Y, Zr, Ag,
and Pd versus Eu with respect to the solar pattern for the most Eu
enriched stars. Clearly this underproduction is a function of Eu
enrichment. \citet{ivans06} recently pointed out that in HD~221170
([Eu/Fe]=0.8) not only the heavy r-process elements, but also the
light r-process elements are in reasonable agreement with the solar
r-process abundance pattern and do not show the pronounced
underproduction of some elements such as Ag and Pd as seen in other
r-process enhanced stars. Given the slopes indicated by the data
displayed in Fig.~\ref{ratios}, one does indeed expect [Sr/Eu],
[Y/Eu], [Zr/Eu], [Pd/Eu], and [Ag/Eu] ratios close to the solar
r-process value for moderately r-process enriched stars around
[Eu/Fe]=0.8.

Recently \citet{honda06} reported seven new elemental abundances in
the metal-poor star HD~122563 and observed an excess of light
neutron-capture elements. This star has ratios of [Ba/Eu]=$-$0.5, 
[Fe/H]=$-$2.7 and [Eu/Fe]=$-$0.5, and the enhancement of light 
neutron-capture elements makes it a candidate for a LEPP enhanced 
metal-poor star.
The abundances of HD~122563 are also shown in Fig.~\ref{ratios}.
They follow nicely the abundance trends found in all the other
metal-poor stars and this consistency makes us believe that
HD~122563 has in fact a significant LEPP contribution.

There are some indications that the observed stable abundance pattern of the main r-process (except for U and Th) extends to the light r-process elements. Fig.~\ref{ratios} shows a flattening of the [X$_{i}$/Eu] vs. [Eu/Fe] slopes for light r-process elements in the most enriched stars with [Eu/Fe]$>$1.3 where the main r-process component dominates. This is most clearly seen for Sr and Y, but Zr, Pd, and Ag are not inconsistent with such a trend. In addition, within the error bars the  [X$_{i}$/Eu] scatter in the [Eu/Fe]$>$1.3 region is small and comparable to the heavier r-process elements. We therefore take the most Eu enriched stars such as CS~22892-052 and CS~31082-001 ([Eu/Fe]$>$+1.0 ) as representatives of a stable, universal r-process component (except for U and Th). The picture that then emerges is that in less Eu enriched stars an additional contribution from the LEPP to the light elements from Sr to Ag becomes visible. We will show below that one then obtains also a LEPP abundance pattern that is fairly consistent from star to star, which again is a hint that our assumptions are justified.

One might argue that there is the possibility of
 a metallicity dependence of the r-process abundance pattern for
 the light r-process elements. Fig.~\ref{metallicity} shows the
 ratio [Sr/Ba] as a function of metallicity. There is no indication
 of a metallicity dependence from these data and at least for
 [Fe/H]$<-$1 they are consistent with a large scatter resulting
 from mixing light and heavy element nucleosynthesis processes
 at low metallicities and a gradual homogenization of the composition
 of the Galaxy as a function of metallicity.

\begin{figure}
\centering
\includegraphics[scale=0.70]{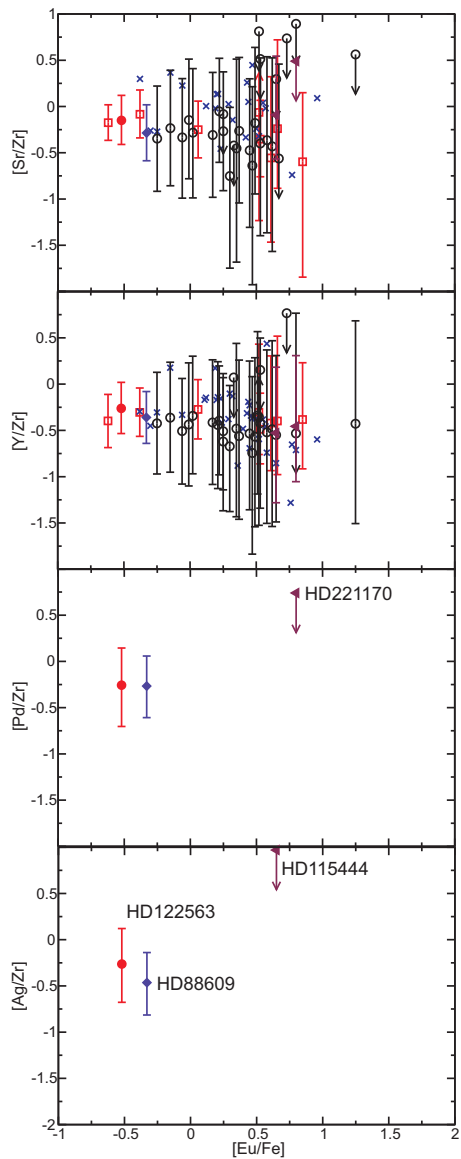}
\caption{Abundance obtained after subtracting a Eu-scaled
CS~31082-001 abundance (main r-process) from metal-poor stars as a
function of [Eu/Fe]. Only stars with [Ba/Eu]$<$0 (except HD~122563)
and [Fe/H]$<-$1 are shown. Abundances represented by crosses were
taken from \cite{burris00}, open squares from \cite{honda04}, open
circles from \cite{christlieb05,barklem05}, filled diamonds from \cite{qian07} 
and filled circles from \cite{honda06}.} 
\label{ratiosWeak}
\end{figure}

In order to find the LEPP pattern in all the metal-poor stars studied, we use CS~31082-001 as an r-process only star, and determine the LEPP abundances in other stars by subtracting its abundance pattern normalized to Eu.
This assumes that all Eu is made in the r-process. The resulting
residual abundances shown in Fig.~\ref{ratiosWeak} were scaled to Zr
so that the patterns can be compared. As  Fig.~\ref{ratiosWeak} shows, the residual
abundances are very consistent for all elements and for all stars
shown. A similar result is obtained by using CS~22892-052 as a
representative of an r-process only star. The scatter of the data in Fig.~\ref{ratiosWeak} measures variations
 within the light element pattern. Clearly the scatter is greatly reduced
 compared to [X/Fe], which has already been shown  for Y, Sr, Zr \citep{Travaglio04}. 
Here we show that the few data on Pd and Ag are consistent with a similar behavior. 
Even though the error bars are large for high [Eu/Fe], they become smaller for 
stars with significant LEPP contribution.
The distribution of the data is consistent with no scatter indicating a consistent 
pattern from star to star for the lighter elements.

We therefore conclude that the LEPP creates a $uniform$
and $unique$ pattern and that with a mixing of a robust r-process, 
the abundance composition of the other metal-poor stars can be obtained.
 For elements from Sr to Ag all weakly Eu-enriched
stars show an overabundance with respect to CS~31082-001, simply
reflecting the extra LEPP component. For elements heavier than Ag,
the LEPP enrichment is less significant and for almost all of the
stars only an upper limit in the abundance can be obtained. 

\begin{figure}
\centering
\includegraphics[scale=0.60]{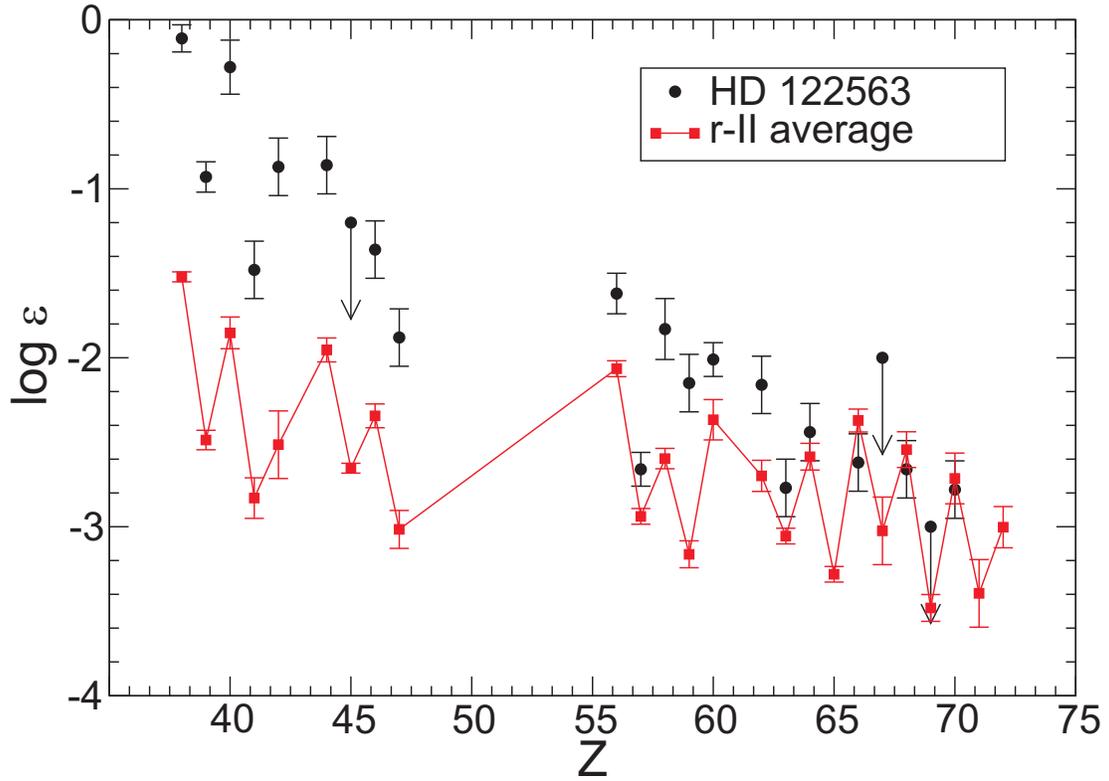}
\caption{Abundance pattern of HD~122563 and scaled abundance pattern
obtained by averaging r-II stars CS~31082-001 and CS~22892-052. The
average r-II stars pattern was normalized to the Eu, Gd, Dy, Er and
Yb HD~122563 abundance.} \label{HD122563}
\end{figure}

To obtain information on the elements that are only weakly produced
in the LEPP one therefore needs to look at the stars with the lowest
[Eu/Fe] where the LEPP most prominently dominates the composition.
We believe that HD~122563 is an example of such a star. Therefore,
having argued for the uniformity and uniqueness of the LEPP
abundance pattern based on a number of stars, we now use HD~122563
to obtain our best estimate of the LEPP abundance pattern. We take
the average of the known r-II stars, and subtract this best
estimate of the r-process from the HD~122563 abundance pattern
assuming that Eu, Gd, Dy, Er and Yb were solely produced in the
r-process (i.e. scaling the main component to those elemental
abundances). Both patterns are shown in Fig.~\ref{HD122563}. The
result is referred to as the {\it stellar LEPP} abundance
pattern in this paper, and it is shown in Fig.~\ref{weak}. It is
noteworthy that we find that some smaller amounts of Ba, Ce, Pr, Nd,
Sm and Eu are still produced by the LEPP.

It would be desirable to obtain a more complete abundance pattern of
more Eu-deficient metal-poor stars that exhibit the same 
(non)correlations as HD~122563. Candidates for such stars are HD~88609 ([Fe/H]=$-$2.93, [Eu/Fe]=$-$0.3), HD~13979 ([Fe/H]=$-$2.26, [Eu/Fe]=$-$0.4) and HD~4306 ([Fe/H]=$-$2.7, [Eu/Fe]=$-$0.6).
However, the abundances of elements from Ru to Ag have not yet been
observed in these stars.

To determine the LEPP contribution to the solar system abundances,
we subtracted the average of the known highly r-process-enhanced
stars from the abundance pattern obtained by subtracting the s- and
p-processes from the solar abundance. Any determination of the solar
r-process abundances suffers from the uncertainties in predicting
the s- and p- process contributions. In particular, uncertainties in
the neutron-capture cross sections and in the solar system
abundances \citep{Arlandini99,Travaglio04} create uncertainties in
the predicted s-process abundances which were taken into account in
the calculations. The solar system abundances were taken from
\citet{anders89} and \citet{lodders03}. The weak component of the
s-process was included using the results of \citet{raiteri92}.
Different models have been used in the past for the main s-process
component, which is the most important one for our study.
\citet{Arlandini99} used the average s-process yield from two AGB
stellar models for a $1.5 M_\odot$ and a $3 M_\odot$ star, both at
metallicity 1/2 $Z_\odot$. On the other hand, \citet{Travaglio04}
followed a Galactic chemical evolution model that used a range of
masses and metallicities and also included intermediate mass star
s-process yields. For elements with $Z\geq50$ both models agree
within the error bars and no major discrepancies are found. While
for elements with $Z\leq37$, \citet{Travaglio04} produce relatively
more s- material than \citet{Arlandini99} due to the additional
contribution in this region by AGB stars of intermediate mass (4 to
8 $M_\odot$), the opposite happens for elements in the range $38\leq
Z\leq51$. The main difference in the resulting solar r-process
contributions are therefore found for Sr, Y and Zr. Using the
s-process contribution from \citet{Arlandini99}, the solar residuals
exhibit smaller amounts of Sr, Y and Zr material than when using the
s-process calculations from \citet{Travaglio04}. Because the
\citet{Travaglio04} Galactic chemical evolution model is more
complete and includes more relevant physics, it is the model that
was used in our study. The effect of the p-process to the elemental
abundance was included by assuming that it solely adds to abundances
of proton-rich isotopes. Only the abundance of Mo and Ru are
significantly modified by the p-process. The elemental LEPP
contribution to the solar system abundances, which we now call the
{\it solar LEPP} pattern, obtained by subtracting the s-, p- and the
r-process (average of r-II stars) from the solar abundances, is
shown in Table~\ref{LEPPabun}. Upper limits of the isotopic solar
LEPP abundances were also obtained. Note that for s-only isotopes
the LEPP abundances can be unambiguously calculated since they do
not have an r-process contribution.

Figure~\ref{weak} compares the solar LEPP pattern with the stellar
LEPP abundance pattern. We find rather good agreement for elements
Y, Sr, Zr, Nb, Mo, Ru and Rh. We therefore propose that the LEPP observed in the abundances of metal poor stars and the process that is responsible for filling in the residual obtained when subtracting from the solar abundances the s-process from \citet{Travaglio04}, the r-process component observed in the most Eu enriched metal poor stars, and the p-process, are the same.
The relative contributions of the LEPP to the solar system abundances
are also in agreement with \citet{ishimaru05}, who found indications
that the LEPP (or weak r-process in their notation) contribution
decreases with atomic number. The element Pd would be somewhat
intermediate between Sr, which is dominated by the LEPP, and Ba,
which is dominated by the main r-process. However, there are also
some discrepancies between the stellar and the solar LEPP patterns.
The solar LEPP abundance of Pd is about a factor of 2 smaller than
the stellar one, but still within 2 $\sigma$ of the error bars.
However, the solar LEPP abundance of Ag is 5 times less than the
stellar LEPP abundance. Since Ag is mainly an r-process element
($\approx$80-86$\%$), it is unlikely that the s-process contribution
is underestimated by more than a factor of 3 to account for the
difference. As mentioned earlier, one possible explanation is the
uncertainty of non-LTE effects in the metal-poor stars abundance
analysis.

\section{Astrophysical conditions}
\label{modelsection}

The second nucleosynthesis process producing the lighter r-process elements postulated in the pioneering work of \citet{wasserburg96} has usually been assumed to be an r-process due to its required production of $^{129}$I. Consequently \citet{qian98} did attempt to model its abundance pattern with a schematic strongly simplified r-process model based on the few observational data that were available at the time.

Traditionally, light s-process products have not been thought 
to be produced at very low metallicities. In particular, the weak 
s-process abundance contribution is
negligible in extremely metal-poor stars with [Fe/H]$\approx-$3 to
account for the abundance of Sr, Y and Zr since the main neutron
source in massive stars, $^{22}$Ne, is of secondary nature. Indeed,
production of $^{22}$Ne derives from the original CNO nuclei, first
converted essentially to $^{14}$N during core H-burning, then
converted to $^{18}$O by $\alpha$ capture at the beginning of core
He-burning and further processed by $\alpha$ capture to $^{22}$Ne.
Neutrons are released by the $^{22}$Ne($\alpha$,n)$^{25}$Mg channel
near core He exhaustion or during the following convective shell
$^{12}$C burning.

Furthermore, the main s-process occurs in low and intermediate mass
stars which have relatively long life spans and cannot thus explain
the observation of metal-poor stars abundances. The isotopic deficiencies of the Galactic chemical evolution s-process model of \citet{Travaglio04} in the s-only isotopes $^{86}$Sr, $^{96}$Mo, $^{100}$Ru, $^{104}$Pd, $^{110}$Cd, $^{116}$Sn, $^{122-124}$Te and $^{128,130}$Xe, which are only produced with abundances of 70-80\% of the solar value, are problematic since those isotopes should
come entirely from the main s-process. Either a third s-process
component has to be included to account for such deficiencies or
there is a problem in their model. For such reason, the possible
existence of a primary s-process contributing to abundances in this
region cannot be excluded. In addition, \citet{froehlich06} and
\citet{wanajo06} have recently suggested that the $\nu$p-process
might contribute to the Y, Sr, and Zr abundances observed in
metal-poor stars. The $\nu$p-process occurs in a proton-rich,
neutrino-driven wind off a proto-neutron star and is therefore a
primary process that could operate in the early Galaxy. While this
process produces primarily neutron-deficient isotopes, it cannot be
excluded observationally as only elemental abundance data are
available for the light r-process elements.

Since we are interested in determining where this second nucleosynthesis process, the LEPP, operates, we explore here the possibility of a neutron capture process being responsible for the LEPP. We do not make any assumptions about the neutron exposure, but
rather determine the necessary conditions to reproduce the newly
obtained LEPP abundance pattern with site-independent full network
calculations. Our goal is to constrain the neutron densities and
temperatures needed for a LEPP process in order to be consistent
with observations. In particular, we want to determine whether the
LEPP is an s- or an r-process (different from the one creating the r-II abundances).

We use a classical approach with a constant neutron exposure of
neutron density $n_{n}$ and duration $\tau$ at a temperature $T$. We
vary conditions from s-process type conditions to r-process type. We
do not use any waiting point or steady flow approximation, but
employ a full reaction network for the abundances of 3224 nuclei
from H to Ta, taking into account neutron capture rates, their
inverse ($\gamma$,n) photo-disintegration rates, and $\beta$-decay
rates with $\beta$-delayed neutron emission. The nuclear reaction
rates were taken from the recent REACLIB compilation, which includes
theoretical reaction rates based on NON-SMOKER statistical model
calculations \citep{rauscher00} with Q-values obtained with the FRDM
\citep{frdm} mass model. Experimental Maxwellian average neutron capture cross sections and their temperature trends were taken from \citet{bao00} when available. Experimental $\beta$-decay rates were used when available \citep{nndc,pfeiffer02}. Theoretical $\beta$-decay rates were taken
from \citet{moeller97} or when available, from calculations
including first forbidden transitions \citep{moeller03}.
Temperature and density dependent $\beta$-decay rates from \citet{takahashi87} were included. The temperature was kept
constant as a function of time. The initial abundance composition
consisted of neutrons and seed nuclei, either $^{56}$Fe, $^{40}$Ca
or a solar distribution seed. The $^{56}$Fe mass fraction used in
the calculation was 1\%. The neutron to seed ratio was chosen such
that $n_{n}$ did not change by more than 5\% during the calculation.
The choice of seed abundance does not affect the final
 abundance pattern, as long as it is below the mass region of interest,
 and therefore none of our conclusions depend on it. Our choice of Fe
 as seed is arbitrary and a purely technical means to create a neutron
 capture flow through the relevant mass region. In particular it does
 not imply the process to be of secondary nature. In general a 
neutron capture process requires some seed. In the case of a 
primary process, this seed had to be created in the same
astrophysical event. An example is 
the $\alpha$-process generating the seed
 for the r-process in the neutrino-driven wind scenario in core
 collapse supernovae.

To quantify which conditions better fit the stellar LEPP abundance
pattern a $\chi^{2}$ function $f(n_{n},T,\tau)$ defined as,
\begin{equation}
f(n_{n},T,\tau)=\sum_{i~\in~LEPP}\left(\frac{Y^{CAL}_{i}-Y^{LEPP}_{i}}{\Delta
Y_{i}} \right)^{2}\textrm{,}
\end{equation}
was used, where $Y^{CAL}_{i}$ is the calculated stable abundance and
$Y^{LEPP}_{i}$ is the desired abundance of element $i$. The closer
the value of $f(n_{n},T,\tau)$ to the number of residuals is, the
better the agreement between the calculated abundance pattern and
the stellar LEPP abundance pattern. Since the size of the
uncertainty of the reference abundances for all elements is
relatively the same, the $\chi^2$ function is not dominated by one
uncertainty and its use is justified.

The duration for the neutron exposure  $\tau$ was chosen to minimize
$f(n_{n},T,\tau)$ for a given set of astrophysical conditions
$n_{n}$ and $T$. In order to do this, we started the calculation for
a given $T$ and $n_n$, and for every time step determined the
abundance pattern that would be produced if the neutrons would be
instantly exhausted at that point and all nuclei would decay back to
stability via $\beta$-decays. The use of a full decay network
including $\beta$-delayed neutron emission for this purpose at every
time step and for all conditions was computationally impracticable.
Beta delayed neutron emission was however included when calculating
the final abundance pattern for the optimum process duration.

\begin{figure}
\begin{center}
\hspace{0.9in}
\includegraphics[scale=0.85]{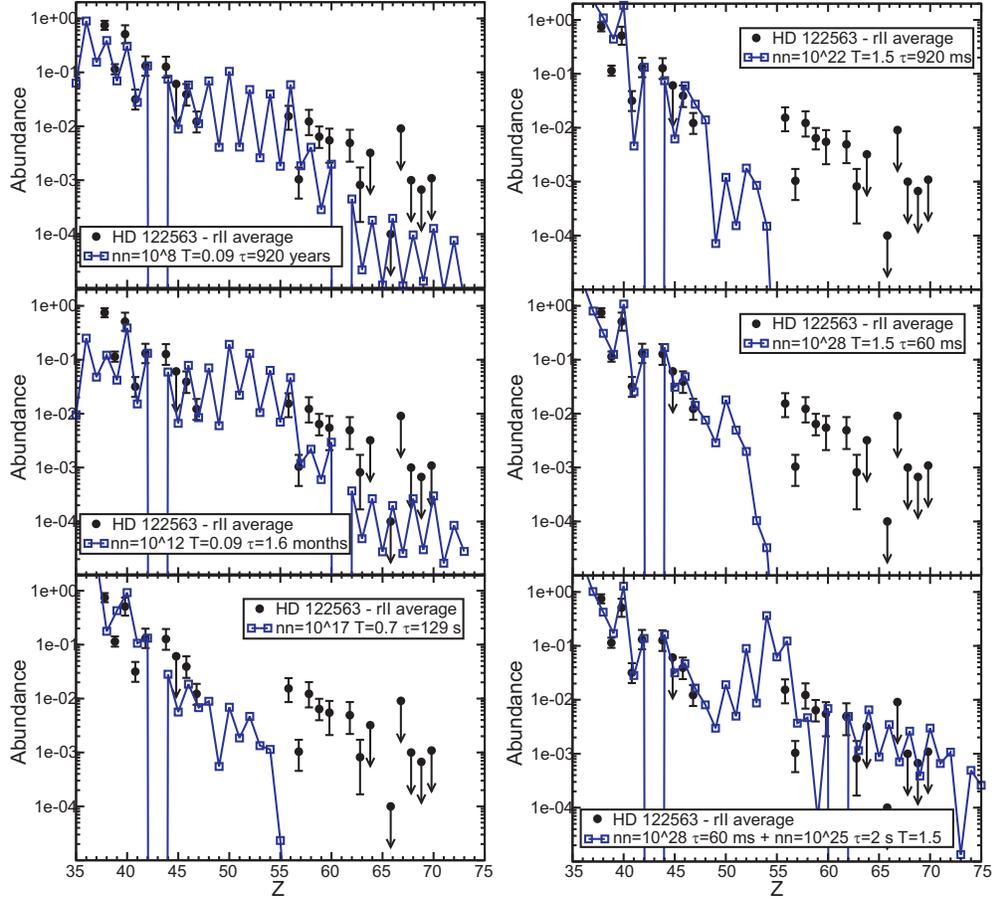}
\end{center}
\caption{Abundances as a function of atomic number normalized to Mo
for different astrophysical conditions and compared with the stellar
LEPP (HD~122563-rII average) pattern. Neutron flux duration was
chosen to better reproduce the stellar LEPP abundances. Neutron
density $n_{n}$ is given in $cm^{-3}$ and temperature $T$ in GK.}
\label{abundancepatterns}
\end{figure}

\begin{figure}
\begin{center}
\hspace{0.9in}
\includegraphics[scale=0.55]{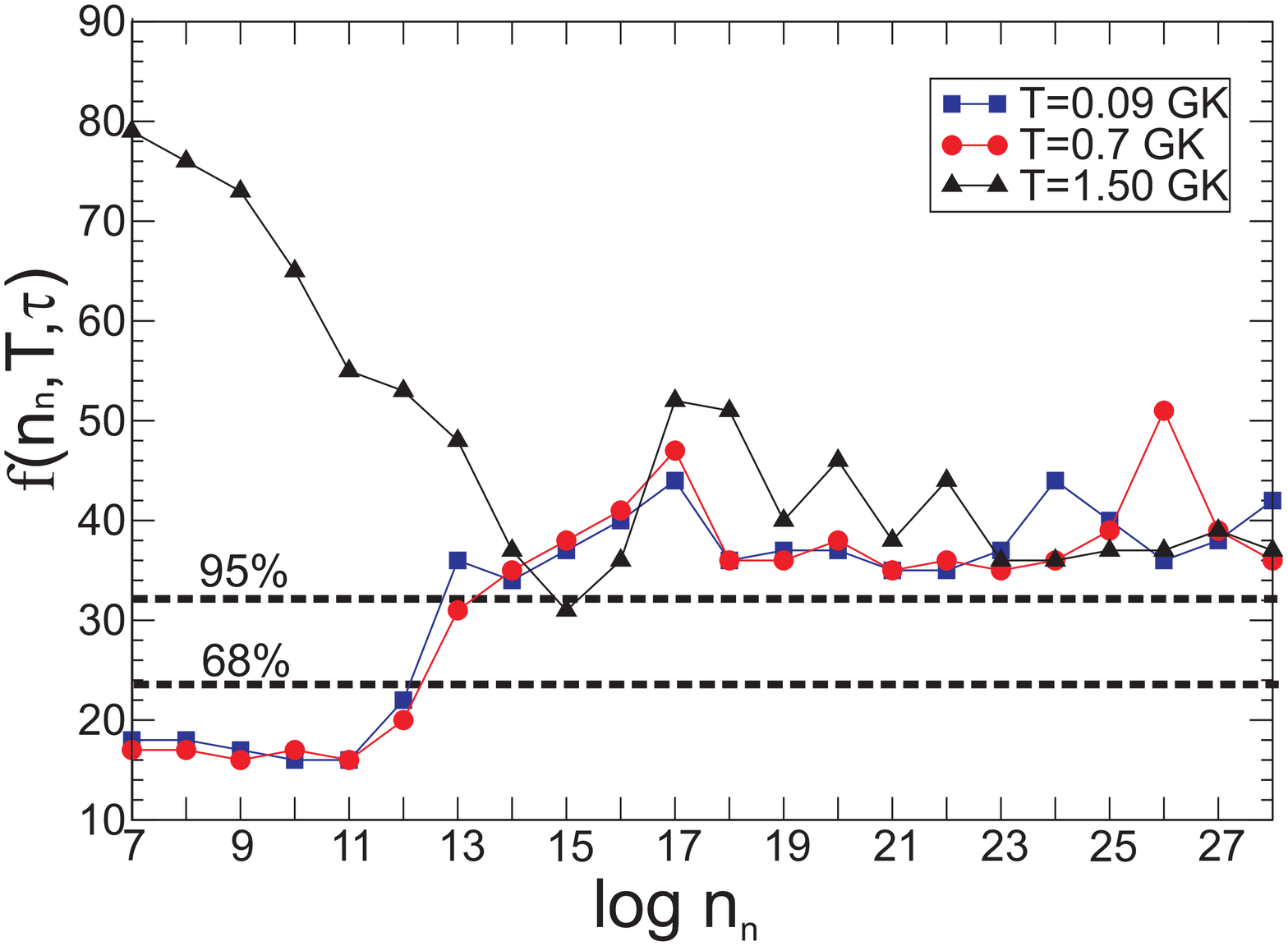}
\caption{$f(n_{n},T,\tau)$ as function of neutron density $n_{n}$
for different temperatures when comparing the results of the network
calculations with the modified HD~122563 abundance pattern.
Confidence intervals for the $\chi^{2}$ distribution are also
shown.} \label{proxHD122563}
\end{center}
\end{figure}

The resulting best abundance patterns for different conditions are
shown in Fig.~\ref{abundancepatterns}. These calculations were
performed with a $^{56}$Fe seed. Using $^{40}$Ca or a solar
abundance distribution as seed did not have a major impact on the
abundance pattern, but only changed the neutron flux duration
$\tau$. We find that the abundances of elements $38\leq Z\leq47$ can
be reasonably reproduced under a variety of different astrophysical
conditions. Even though low neutron densities $n_{n}\approx10^{8}$
cm$^{-3}$ and high neutron densities $n_{n}\approx10^{28}$ cm$^{-3}$
can fit the LEPP pattern best in this region, other neutron
densities can reproduce the pattern within a factor of 3 for every
element. However, if heavier elements ($Z\geq56$), even in a
relatively low amount such as in the LEPP, also have to be created,
low neutron densities are favored to reproduce the desired
abundances. This is illustrated in Figure~\ref{proxHD122563}, which
displays $f(n_{n},T,\tau)$ for different astrophysical conditions.
Neutron-capture processes with a low neutron density,
$n_{n}\leq10^{13}$ cm$^{-3}$, reproduce the residual abundance
pattern better. Higher neutron densities fail to reproduce the
abundance pattern since it is not possible (within the model) to
create the correct abundances for $38\leq Z\leq47$ and sufficient
amounts of $Z\geq56$ material. The dependence on temperature is
small; only for low neutron densities can a high temperature be
excluded.

The average atomic number of the created nuclei increases as a
function of time as neutron captures are followed by $\beta$-decays.
As the material becomes heavier, some of it reaches the region
$38\leq Z\leq47$ and the desired abundance pattern may be
reproduced. As more and more material increases its atomic number,
the abundance in the region $38\leq Z\leq47$ decreases and the
abundance of $Z\geq56$ starts to increase.  To satisfactorily
reproduce the residual abundance pattern most of the abundance has
to go into $38\leq Z\leq47$. For $Z\geq56$, the amount of created
material has to be about one order of magnitude less than the
average abundance of the light elements. The neutron shell closure
$N=82$ is a bottleneck where abundances accumulate. In order to
produce sufficient  $38\leq Z\leq47$ abundances the neutron flux has
to be exhausted while most of the material is passing through the
$N=82$ bottleneck. For processes with a large neutron density, the
abundance peak occurs around $50\leq Z\leq56$.  In order to produce
elements $56\leq Z\leq62$, enough material has to leak out of the
bottleneck. Because of the relatively long time for that to occur,
the abundance of elements $Z=47$ and $48$ already decreases before
enough $56\leq Z\leq62$ material is created. For processes with a
relatively small neutron density, the shell closure produces
progenitor bottleneck abundances in the region $56\leq Z\leq60$ and
therefore the required amount of heavy material can still be
obtained.

Even though low neutron density scenarios produce a higher amount of
heavy material that is in agreement with the solar LEPP abundances,
the stellar LEPP abundance pattern is not completely reproduced for
$Z\geq56$ as shown in Fig.~\ref{abundancepatterns}.  While Ba seems
to be always overproduced, Pr and Sm are underproduced by the
network model using low neutron density scenarios. The neutron flux
duration necessary to obtain a reasonable fit under the lowest
neutron density ($\tau \approx 1170$ years for $n_{n}=10^{8}$
cm$^{-3}$) far exceeds what is expected by present nucleosynthesis
calculations in massive stars
\citep{woosley02,rauscher02,chieffi04}. More favorable conditions
would require higher $n_{n}$. The choice $n_{n}=10^{13}$ cm$^{-3}$
implies $\tau \approx 5$ days. A primary neutron source would also
be necessary to obtain a primary-LEPP mechanism, either by the
elusive channel $^{12}$C+$^{12}$C$\rightarrow$n+$^{23}$Mg, or by
more sophisticated situations in which convective shell C-burning
layers merge with hotter inner regions suffering Ne-shell and
O-shell burning in the most advanced phases. These developments are
outside the scope of the present analysis.

Even though in the high end of the $n_n\le 10^{13}$ cm$^{-3}$ range some amount of $^{129}$I is made (at a neutron density $n_n \approx 10^{11}$ cm$^{-3}$ the neutron capture on $^{127}$Te becomes comparable to the beta decay halflive $\approx10$ h and therefore a subsequent neutron capture on the stable $^{128}$Te and the $^{129}$Te beta decay creates it), in our calculations not enough $^{127}$I is produced to explain the meteoritic ratios. It is however interesting that for low neutron densities, $n_{n}\leq10^{11}$ cm$^{-3}$, the derived solar LEPP isotopic abundances are in agreement with the missing s-process abundances in the region Mo to Xe predicted by the Galactic chemical evolution model by \citet{Travaglio04} that includes the yields of all AGB stars according to their lifetimes and production at various metallicities. In particular, s-only isotopes in the region from Mo to Xe are within 20 to 30$\%$ of their solar abundances. For higher neutron densities, the LEPP isotopic distribution shifts progressively towards an r-process behavior.

The disadvantage of using a site-independent 
model is that the reaction network calculations may be over 
simplistic and some important features can be left out. 
R-process scenarios such as the neutrino driven wind 
in supernovae and supernova fallback 
\citep{WWH92,TWJ94,Wan01,Tho01,Far06,fryer06} 
have neutron densities that dramatically evolve with time. 
By keeping a single constant neutron density the effect of 
such change cannot be correctly reproduced. For such a reason 
we also performed test calculations to explore whether separate 
r-process components for the $38\leq Z\leq47$ and the 
$56\leq Z\leq62$ regions could reproduce the LEPP abundance 
pattern. Fig.~\ref{abundancepatterns} shows the abundance 
pattern when using components $n_{n}=10^{28}$ cm$^{-3}$, 
$\tau=60$ ms and $n_{n}=10^{25}$ cm$^{-3}$, $\tau=2$ s 
at $T_{9}=1.5$. Although the stellar LEPP is not completely 
reproduced for the heavy elements, the abundance pattern 
is reproduced to better than an order of magnitude with the exception 
of Ba and Pr. Even though the choice of components is not unique, 
a lower neutron density limit of $n_{n}=10^{24}$ cm$^{-3}$ 
was found preferable to reproduce light and heavy elements 
without overproducing Pd, Ag, though some overproduction of 
Ba cannot be avoided. In this case, no s-process production is
possible.  Site-dependent 
calculations should be performed in the future to compare
the observed LEPP abundances with predictions from various
realistic scenarios.

\section{Conclusions}
\label{conclusions}
\begin{figure}
\centering
\includegraphics[scale=0.60]{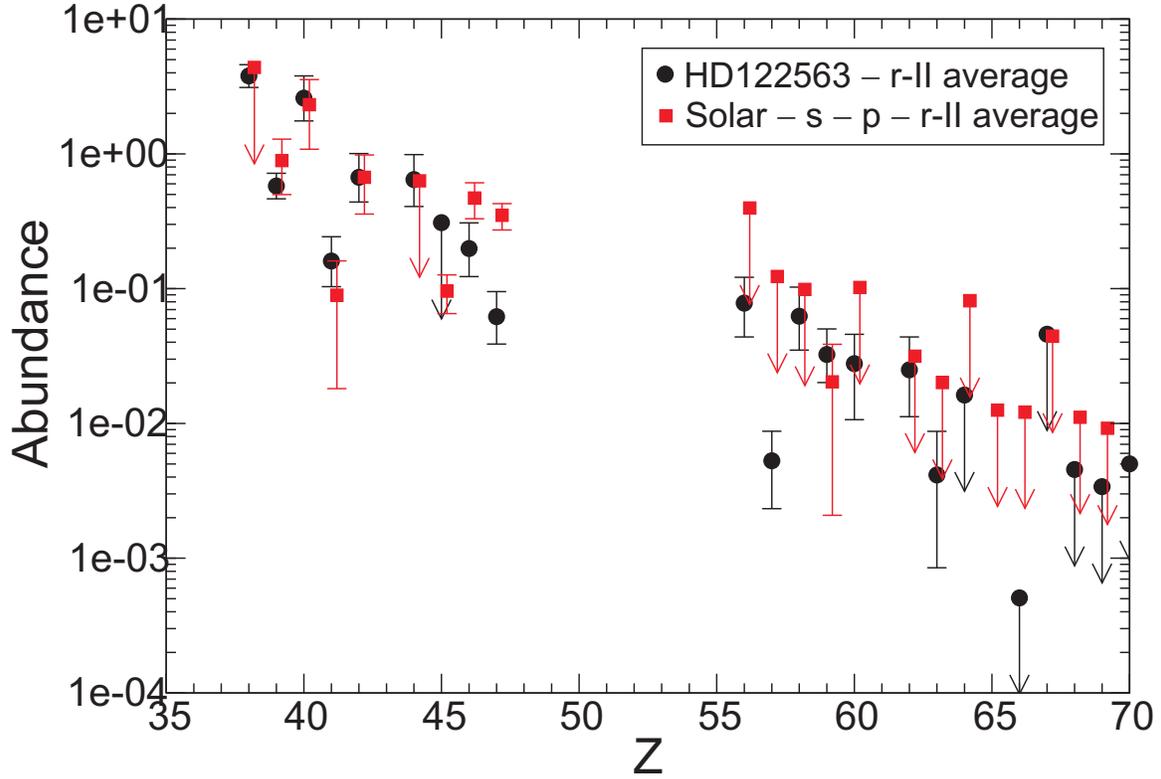}
\caption{Abundance pattern created by the LEPP.  Pattern represented
by filled squares was created by subtracting the scaled average of
CS~31082-001 and CS~22892-052, the s- and p- process contributions from
the solar abundance (solar LEPP). Pattern represented by filled circles 
was obtained 
by subtracting the scaled average of
CS~31082-001 and CS~22892-052 from HD~122563 (stellar LEPP) and scaling 
it to the
solar LEPP Mo abundance. Read text for explanation.}
\label{weak}
\end{figure}

We have shown that the elemental abundances of metal-poor halo stars
exhibit a non-correlation between [X/Eu] and [Eu/Fe] for Y, Sr, Zr,
Pd, and Ag. The same behavior had been found before for Y and Zr in
a few stars in M15 \citep{kaori06}. This provides new further
evidence for the existence of a primary LEPP process that
contributes together with the r-process, the weak s-process, the
main s-process, and the p-process to the nucleosynthesis not only of
Y, Sr, and Zr, but, as we show here, for most elements in the Sr-Ag
range. We also find that a very small contribution to still heavier
elements up to Eu is likely. Based on our results we were then able
to show that the LEPP produces a uniform and unique abundance
pattern, shown in Fig.~\ref{weak}, and that together with the pattern observed in Eu-enriched stars (r-process rich), are able to explain the abundances of all metal-poor stars considered.

Metal-poor stars with very weak Eu enhancement play an essential
role in constraining the LEPP as they have the smallest contribution
from the r-process. A prime example is HD~122563, for which a wide
range of elemental abundances are observed. We are therefore able to
explain the abundance observations in HD~122563 and HD~221170 that
previously had been identified as ``anomalies", together with the
abundances observed in other metal-poor halo stars with a consistent
picture of mixed contributions from the r-process and the LEPP.
In addition, it was found that the LEPP 
contributes significantly to the solar system abundances 
based on the use of the \citet{Travaglio04} s-process model. 
While we consider this model to be the best available it 
should be noted that the use of the simpler s-process 
model by \citet{Arlandini99}, for example, would have 
led to a significantly reduced solar system contribution 
of the LEPP. However, only when using the \citet{Travaglio04} 
s-process model do the solar and the stellar LEPP 
abundance patterns agree.

Since the astrophysical conditions that would create the LEPP
abundance pattern are not known, full reaction network calculations
were performed in a heuristic way assuming different neutron capture
process conditions. A variety of different neutron densities from
s-process to r-process like were found to reproduce the abundance
pattern between Sr and Ag within the observational uncertainties.
However, intermediate neutron densities in between typical s- or
r-process conditions seem to be excluded. Using a single component
to reproduce the LEPP pattern only neutron densities $n_{n}\leq10^{13}$
cm$^{-3}$ seem to create enough Ba to Sm material (which actually consist
of quite small contributions to solar) that is consistent within
an order of magnitude with the abundances inferred for HD~122563.
These low neutron densities correspond to densities found in the
s-process, or not so far from it.
A LEPP characterized by neutron densities of $n_{n}\leq10^{13}$ cm$^{-3}$
 then not only addresses the the problem of explaining metal poor star abundance patterns,
 but also the problem of the underproduction of some s-only isotopes in the s-process
 galactic chemical evolution model of \citet{Travaglio04}.

Multiple nucleosynthesis processes are also required to explain the early solar system $^{129}$I/$^{127}$I 
and $^{182}$Hf/$^{180}$Hf ratios inferred from meteorites \citep{qian03}. 
As $^{129}$I and $^{182}$Hf are radioactive nuclei with different 
half-lives, the detected abundance ratios imply different chemical evolution 
histories for $^{129}$I and $^{182}$Hf, both thought to be produced 
in the r-process. If low neutron density scenarios are responsible 
for the LEPP, the A=130 abundance peak could be attributed to 
the main r-process component (as observed in r-II metal-poor stars). 
In this case, most of $^{129}$I and $^{182}$Hf would be produced in the same 
r-process events, which could not explain the meteoritic data. 
It should be noted however that \citet{meyer00} have questioned 
the pure r-process origin of $^{182}$Hf. An alternative scenario that would
satisfy the meteoritic constraints would be that yet another process
is responsible for the origin of the $A=130$ abundance peak. This would
require that the $A=130$ production, or at least the production of 
$^{129}$I, be largely avoided in the main r-process that 
is known to produce the heavy elements from Ba and beyond. Similarly, 
the hypothetical additional process that is responsible for the synthesis of $^{129}$I
would have to provide negligible contributions to Ba. It would 
have to be demonstrated in realistic model calculations that
both requirements can be achieved. Recent studies based on the 
classical r-process model indicate that this might be difficult
given the know nuclear physics around the $N=82$ shell closure \citep{kratz06}.

The astrophysical scenarios involving  neutron densities $n_{n}\leq10^{13}$ cm$^{-3}$ do not correspond to the traditional
weak or main s-process because the nucleosynthesis occurs in very
low metallicity stars and the required neutron flux duration is too
long compared to what is expected in those scenarios. A particular
challenge is to find a stellar scenario with low neutron densities
during a long period of time occurring in low metallicity stars
strong enough to produce elements up to Eu. Since it is hard to
envision such a scenario, possibilities other than low neutron capture processes should also be considered 
to explain the observed LEPP abundances.

While it is not possible to reproduce the entire LEPP abundance pattern at high neutron densities with a single neutron exposure, we showed that in principle a multi-component exposure with neutron densities $n_{n}\geq10^{24}$ cm$^{-3}$
could reproduce the observed abundances. In such a model
the LEPP could synthesize $^{129}$I explaining the meteoritic data,
though the overproduction of Ba is difficult to avoid. 
A high neutron density LEPP would however require
that the solar abundance residual (our solar LEPP abundance pattern)
cannot, or at least not entirely, be explained with LEPP anymore, as it
contains the s-only nuclei underproduced in \citet{Travaglio04}.
In this case one would have to conclude that the LEPP
contributes at most a small amount to the solar abundances, and that 
an unknown additional s-process component is required to explain the solar
abundances. Moreover, the agreement between our solar LEPP 
pattern and the observed LEPP component in metal poor stars 
pointed out in this work would than have to be considered coincidental.

The s-process contribution to solar $^{96}$Mo is only 78\%. Since
that isotope is shielded by $^{96}$Ru against an rp-process far from
stability, one might argue that such a process is
possibly excluded as an explanation for the LEPP. Nevertheless, a
nucleosynthesis process on the proton-rich side running closer to
stability such as the recently proposed $\nu$p-process should be
considered. Besides proton captures, the $\nu$p-process includes
neutron induced reactions and therefore has a path closer to
stability producing isotopes such as $^{96}$Mo. In addition,
\citet{wanajo06} have shown that the $\nu$p-process under some
conditions can produce enough material up to Eu. Further studies
should also consider this process a candidate for the production of
LEPP abundances.

The parameter study in the present work is a first step toward
determining the astronomical site responsible for creating the
abundance of material not created in the r-process in metal-poor
stars. More observational data, particularly for r-process poor
stars and for more elements below Ba, would certainly be important
for further progress. It would also be desirable to identify actual
sites that could be responsible for the LEPP and perform site
specific calculations to reproduce our derived LEPP abundance
pattern.

After submission of this paper, \citet{qian07} presented a
refinement of their phenomenological model that is based on
similar observational constraints as presented here. Their model is
based on the observed abundances in HD~122563 (for their ``L"
component) and CS~22892-052 (for their ``H" component). Tough their
adopted patterns are slightly different, their conclusion that such
a two component model can explain currently available metal poor
star abundance patterns is in agreement with this work. Our results
concerning the implications of the production of some A$\ge$130
nuclei in the LEPP, the likely nature of the LEPP and its potential
relevance for the s-process are not affected.

\acknowledgments We thank I.I. Ivans, J. E. Lawler, C. Sneden and K.
Lodders for helpful discussions, and F.-K. Thielemann for providing
the reaction network solver. This work has been supported in part by
NSF grants PHY 02-16783 (Joint Institute for Nuclear Astrophysics)
and PHY 01-10253. Support was also provided by NSF grants
AST03-07279 (J.J.C) and AST04-06784 (T.C.B), by the Virtuelles
Institut f{\"u}r Struktur der Kerne und Nukleare Astrophysik under
HGF grant VH-VI-061, by the Univ. Mainz-GSI F+E Vertrag under grant
MZ/KLK (K.-L.K, B.P. and K.F), and by the Italian MIUR-FIRB Project
``Astrophysical Origin of the Heavy Elements Beyond Fe'' (R.G.).

\clearpage



\begin{table}[p]
\caption{LEPP contribution to the solar system total abundance
(Solar LEPP as defined in the text).}
\begin{center}
\begin{tabular}{|c|c|}\hline
Element & $\%$ \\
\hline \hline
Sr & $\leq$19 \\
Y & 19(8) \\
Zr & 20(11) \\
Nb & 13(10) \\
Mo & 26(12) \\
Ru & $\leq$34 \\
Rh & 28(9) \\
Pd & 34(10) \\
Ag & 72(16) \\
Ba & $\leq$9 \\
La & $\leq$28 \\
Ce & $\leq$8 \\
Pr & 12(11) \\
Nd & $\leq$12 \\
Sm & $\leq$12 \\
Eu & $\leq$21 \\
Gd & $\leq$25 \\
\hline
\end{tabular}
\end{center}
\label{LEPPabun}
\end{table}


\clearpage

\begin{figure}
\includegraphics[scale=0.60]{f2.eps}
\caption{Abundance ratio [Sr/Eu] of metal-poor stars as a function
of metallicity [Fe/H]. Only stars with [Ba/Eu]$<$0 (except
HD~122563) and [Fe/H]$<-$1 are shown. Abundances represented by
crosses were taken from \cite{burris00}, open squares from
\cite{honda04}, open circles from \cite{christlieb05,barklem05} and
filled circles from \cite{honda06}.} \label{metallicity}
\end{figure}

\clearpage

\clearpage

\clearpage

\clearpage

\clearpage

\clearpage


\end{document}